\documentstyle[amssymb,aps,graphicx,subeqnarray]{revtex}

\draft

\begin{document}

\twocolumn[\hsize\textwidth\columnwidth\hsize\csname @twocolumnfalse\endcsname
\title{Numerical computation of one-photon mazer resonances for arbitrary field modes}

\author{T. Bastin$^{1}$ and E. Solano$^{2,3}$}
\address{$^{1}$Institut de Physique Nucl\'eaire Exp\'erimentale, Universit\'e
  de Li\`ege au Sart Tilman, B\^at. B15, B - 4000 Li\`ege, Belgique\\
$^{2}$Instituto de F\'{\i}sica, Universidade Federal do Rio de Janeiro, Caixa Postal
68528, 21945-970 Rio de Janeiro, RJ, Brazil\\
$^{3}$Secci\'{o}n F\'{\i}sica, Departamento de Ciencias, Pontificia
Universidad Cat\'{o}lica del Per\'{u}, Apartado 1761, Lima, Peru}

\date{22 July 1999}

\maketitle

\begin{abstract}
We present a novel approach for solving numerically
one-dimensional scattering problems and apply it for computing the
emission probability of an ultracold atom interacting with an
arbitrary field mode of a high-$Q$ cavity. Our method is
efficient, stable and succeeds when other numerical integration
methods fail. It also replaces and improves advantageously the WKB
approximation. The cases of sinusoidal, $\mathrm{sech}^2$ and
Gaussian field modes are studied and compared. Divergences with
previous works, where WKB was used, are found.
\end{abstract}


\vskip2pc]

\section{Introduction}
\label{Introduction} Some recent works have been devoted to the
interaction of ultracold atoms with microwave cavities
\cite{Scu96,Mey97,Lof97,Sch97,Ret98,Zha99}. These studies treated
the interaction between an incident atom in an excited state and a
cavity field containing $n$ photons, taking the quantum mechanical
CM motion of the atom into account. This interaction leads to a
new kind of induced emission intimately associated with the
quantization of the CM motion, named the mazer action
\cite{Scu96}.

After interaction with the cavity, the atom can be found
transmitted in the excited state or in the lower state, or
reflected as well. It has been shown \cite{Scu96} that these event
probability amplitudes (denoted respectively $T_{a,n}$,
$T_{b,n+1}$, $R_{a,n}$ and $R_{b,n+1}$) are given by
\begin{equation}
  \label{TRinExpr}
  \renewcommand{\arraystretch}{1.4}
  \begin{array}{cc}
    T_{a,n}=\frac{1}{2}( t_{+n}+t_{-n} ) \qquad &
    T_{b,n+1}=\frac{1}{2}( t_{+n}-t_{-n} ) \\
    R_{a,n}=\frac{1}{2}( r_{+n}+r_{-n} ) \qquad &
    R_{b,n+1}=\frac{1}{2}( r_{+n}-r_{-n} )
  \end{array}
  \renewcommand{\arraystretch}{1.0}
\end{equation}
where $t_{\pm n}$ and $r_{\pm n}$ are, respectively, the transmission
and reflection amplitudes of the elementary scattering process of
the atom incident upon the potential
\begin{equation}
  V_{n}^{\pm}(x)=\pm \hbar g \sqrt{n+1}\:u(x)
\end{equation}
where $x$ is the atom traveling direction, $g$ is the atom~--
field coupling strength and $u(x)$ is the cavity field mode
function.

The induced emission probability of the atom interacting with the cavity field is then given by
\begin{equation}
  \label{Pem}
  P_{em}=|T_{b,n+1}|^2+|R_{b,n+1}|^2
\end{equation}

All these probabilities are strongly dependent on the cavity mode
profile. Their calculation needs to solve the one-dimensional
time-independent Schr\"odinger equation. L\"{o}ffler \textit{et~al.}~\cite{Lof97} have calculated them as a function of the interaction
length $\kappa_{n}L$ ($L$ is the cavity length and
$\kappa_{n}=\kappa \sqrt[4]{n+1}$ with $\kappa=\sqrt{2mg/\hbar}$,
and $m$ the atomic mass) for various cavity mode functions: mesa,
$\textrm{sech}^2$ and sinusoidal modes. For the two first modes,
analytical results of the probability $P_{em}(\kappa_{n}L)$ have
been given. For the sinusoidal modes, detailed Wentzel-Kramers-Brillouin solutions have been presented. Nevertheless,
Retamal \textit{et~al.}~\cite{Ret98} have shown that the WKB
approximation may lead to inaccurate predictions for actual
interaction and cavity parameters.

The consideration of actual interaction and cavity parameters is a difficult
numerical task as they can correspond to very large values of
$\kappa_{n}L$. When no analytical solutions of the Schr\"odinger
equation are available, usual numerical integration methods do not
converge rapidly or do not converge at all.

In this paper, we present a novel approach for solving the
one-dimensional time-independent Schr\"odinger equation. Our
method is described in Sec.~\ref{MethodDescription}. It may be
applied for any potential function. It is efficient, stable and
succeeds when other numerical integration methods fail. It may
replace and improve advantageously the WKB approach. Details of
the implementation of our method and validity tests are presented
in Sec.~\ref{MethodImplementation}.

We apply our method for calculating some new $P_{em}(\kappa_{n}L)$
curves. In particular, the Gaussian potential has been considered,
thinking in open cavities in the microwave or optical field
regime. These results are presented in Sec.~\ref{NewResults}.

\section{Description of the method}
\label{MethodDescription}
The one-dimensional time-independent Schr\"odinger equation can be
written in atomic units in the form
\begin{equation}
  \label{SchrEq}
  \left( \frac{1}{2}\frac{d^2}{dx^2}+(E-V(x))\right)\phi(x)=0
\end{equation}

We assume here that the potential $V(x)$ has non zero values only in a
given region of the $x$ axis, say $\left[x_a,x_b\right]$. This region
is divided into $J$ grid points, denoted $x_1,\ldots,x_J$ such that
$x_a=x_1<x_2<\cdots<x_{J-1}<x_J=x_b$. Let us note $I_j$ ($0<j<J$) the
region $\left[x_j,x_{j+1}\right]$, $I_0$ $\left]-\infty,x_1\right]$,
and $I_J$ $\left[x_J,+\infty\right[$.

The potential $V(x)$ is approximated in each region $I_j$ by a
straight line connecting $V(x_j)$ and $V(x_{j+1})$. The approximated
potential is noted $V_{approx}(x)$. Then Schr\"odinger equation takes the simple form in
$I_j$ ($0 \leq j \leq J$):
\begin{equation}
  \label{ApproxSchrEq}
  \left( \frac{d^2}{dx^2}+(a_j+b_{j}x)\right)\phi_{j}(x)=0
\end{equation}
with $a_0=a_J=2E$, $b_0=b_J=0$, and for $0<j<J$
\begin{equation}
  \label{ajbjDef}
  \left\{ \begin{array}{l}
    a_j=2 \left(E-\frac{x_{j+1}V_{j}-x_{j}V_{j+1}}{x_{j+1}-x_{j}}\right)\\
    b_j=-2 \frac{V_{j+1}-V_{j}}{x_{j+1}-x_{j}} \quad \textrm{with}
    \quad V_i=V(x_i),i=j,j+1
  \end{array} \right.
\end{equation}

The most general solutions of Eq.~(\ref{ApproxSchrEq}) are given by
\begin{equation}
  \label{phij}
  \phi_{j}(x)=C_{j}f_{j}^{+}(x)+D_{j}f_{j}^{-}(x)
\end{equation}
where $C_j$ and $D_j$ are two complex constants and $f_{j}^{+}(x)$
and $f_{j}^{-}(x)$ 2 functions depending on the $a_j$ and $b_j$
values. These functions are given in Table~\ref{FunctionsTable}.

As shown in Table~\ref{FunctionsTable}, the functions
$f_{j}^{+}(x)$ and $f_{j}^{-}(x)$ depend on the sign of
$a_j+b_{j}x$. To avoid a change of this sign in a given region
$I_j$, the set $x_1,\ldots,x_J$ must contain the roots of the
equation $V(x)=E$. Thus, the condition $a_j+b_{j}x>0$ (resp.
$a_j+b_{j}x<0$) is equivalent to that $E>V(x)$ (resp. $E<V(x)$)
for $x \in I_j$.

The complex constants $C_j$ and $D_j$ in Eq.~(\ref{phij}) are
determined by use of the wavefunction asymptotic behaviour
knowledge ($(C_0,D_0)$ or $(C_J,D_J)$ are supposed to be known)
and of the set of conditions imposing the continuity of the
wavefunction and its derivative along the $x$ axis~:
\begin{equation}
  \label{ContinuityConditions}
  \left\{ \begin{array}{l} \phi_{j}(x_j)=\phi_{j-1}(x_j) \\
  \phi'_{j}(x_j)=\phi'_{j-1}(x_j) \end{array} \right. \quad (0 < j \leq J).
\end{equation}

If $(C_0,D_0)$ are given, Eq.~(\ref{ContinuityConditions}) may be
written as
\begin{equation}
  \label{ContinuityConditionsA}
  \left( \begin{array}{l} C_{j} \\ D_{j} \end{array} \right) = A_{j}(x_j)
  \left( \begin{array}{l} C_{j-1} \\ D_{j-1} \end{array} \right)
  \quad \textrm{for} \: 0 < j \leq J,
\end{equation}
with
\begin{eqnarray}
  A_j & = & \frac{1}{f_{j}^{+}g_{j}^{-}-f_{j}^{-}g_{j}^{+}} \times \nonumber\\
      &   & {}\left( \begin{array}{cc}
                f_{j-1}^{+}g_{j}^{-}-f_{j}^{-}g_{j-1}^{+} &
                f_{j-1}^{-}g_{j}^{-}-f_{j}^{-}g_{j-1}^{-} \\
                f_{j}^{+}g_{j-1}^{+}-f_{j-1}^{+}g_{j}^{+} &
                f_{j}^{+}g_{j-1}^{-}-f_{j-1}^{-}g_{j}^{+}
              \end{array} \right)
\end{eqnarray}
and
\begin{equation}
  g_i^{\pm}(x)=\frac{df_i^{\pm}}{dx} \, , \,i=j,j-1.
\end{equation}

If $(C_J,D_J)$ are known, Eq.~(\ref{ContinuityConditions}) should
be written
\begin{equation}
  \label{ContinuityConditionsB}
  \left( \begin{array}{l} C_{j-1} \\ D_{j-1} \end{array} \right) = B_{j}(x_j)
  \left( \begin{array}{l} C_{j} \\ D_{j} \end{array} \right)
  \quad \textrm{for} \: J \geq j > 0,
\end{equation}
with
\begin{eqnarray}
  B_j & = & \frac{1}{f_{j-1}^{+}g_{j-1}^{-}-f_{j-1}^{-}g_{j-1}^{+}} \times \nonumber\\
      &   & {}\left( \begin{array}{cc}
                 f_{j}^{+}g_{j-1}^{-}-f_{j-1}^{-}g_{j}^{+} &
                 f_{j}^{-}g_{j-1}^{-}-f_{j-1}^{-}g_{j}^{-} \\
                 f_{j-1}^{+}g_{j}^{+}-f_{j}^{+}g_{j-1}^{+} &
                 f_{j-1}^{+}g_{j}^{-}-f_{j}^{-}g_{j-1}^{+}
               \end{array} \right).
\end{eqnarray}

The knowledge of $\phi_{j}(x)$ in each region $I_j$ defines an
approximated wavefunction of the Schr\"odinger equation
(\ref{SchrEq}). The grid point number $J$ fixes the accuracy of
the method. The higher $J$ is, the more accurate is the
approximated potential $V_{approx}(x)$ and the better is the
approximated wavefunction. The actual number $J$ to be considered
for a given accuracy depends on various parameters (potential
form, energy $E$, size of the region $\left[x_a,x_b\right]$). We
show in Sec.~\ref{MethodImplementation} and \ref{NewResults} that
a few hundreds are typical numbers.

The wavefunctions yielded by our method are not divergent at the
classical turning points (where $E = V(x)$), on the contrary of
the WKB method. Although some Bessel functions of
Table~\ref{FunctionsTable} are divergent at these points (as
$z_j(x) \equiv a_j + b_j x = 0$), all the functions $f_{j}^{\pm}(x)$ and their
derivative $g_{j}^{\pm}(x)$ admit finite limits there (see Table~\ref{Limits}).

The transmission and reflection complex amplitudes (denoted
respectively $t$ and $r$) of a particle incident upon a potential
$V(x)$ may be calculated as well (and consequently the induced
emission probability (\ref{Pem})). If we consider the outgoing
wavefunction $\phi_{J}(x)=e^{ikx}$ with $k=\sqrt{2E}$, we can
calculate the corresponding incoming wavefunction
$\phi_{0}(x)=C_0\cos(kx)+D_0\sin(kx)$ by use of
relations~(\ref{ContinuityConditionsB}). We then have
\begin{equation}
  \label{tAndr}
  t=\frac{2}{C_0-iD_0} \quad , \quad r=\frac{C_0+iD_0}{C_0-iD_0}
\end{equation}

\section{Method implementation}
\label{MethodImplementation}

We have implemented the method described above on a PC Pentium
based computer. Bessel functions of
Table~\ref{FunctionsTable} have been coded using algorithms given
by Zhang and Jin~\cite{Zha96}. The evaluation of the Mathematica
$^{\circledR}$
Bessel functions has been discarded due
to poor performances (thousand times slower than direct
computation). For usual cavity parameters, $C_0$ and $D_0$
coefficients of Eq.~(\ref{tAndr}) can rapidly become very large.
To avoid computation overflow errors, we have adopted a custom number
representation. Our internal range of values was
9.9~E~$\pm$~($\sim$9~E~18) with 15 significants.

\begin{figure}

\begin{center}
\noindent\fbox{\includegraphics[width=8cm, trim= 30 400 40 120,
keepaspectratio]{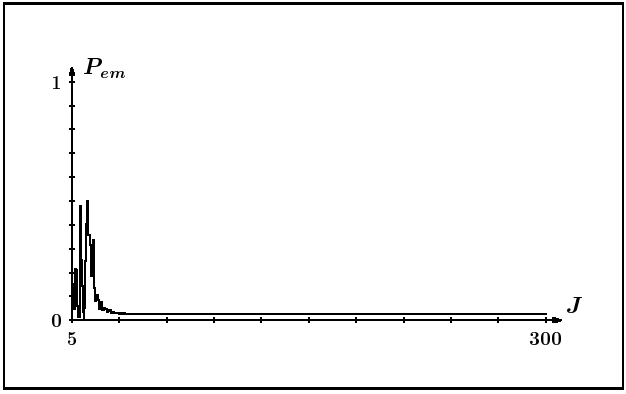}}
\end{center}

\caption{The induced emission probability $P_{em}$ as a function
of the grid point number $J$ for $\kappa_nL=10$
($k/\kappa_n=0.01$, $\textrm{sech}^2$ cavity mode profile).}
  \label{FigurePem_JSech2}
\end{figure}

As a first test of our method, we checked its ability to converge
when the grid point number $J$ increases. We have calculated the
induced emission probability $P_{em}$ of an atom interacting with
a sech${}^2$ mode profile cavity ($u(x)=\textrm{sech}^2(x/L)$) as
a function of the number $J$ and for a fixed interaction length
$\kappa_{n}L$. Fig.~\ref{FigurePem_JSech2} presents the curve
obtained in the case of $k/\kappa_n=0.01$ and $\kappa_nL=10$. This
curve shows very clearly that $P_{em}$ converges to a given value
as $J$ increases. Also, in that case, a $J$ number of 200 is
sufficient for predictions to the accuracy of the graphics. For
this calculation, the region $\left[x_a,x_b\right]$ was limited to
16 times the cavity length $L$ ($\sim 8$ times the FWHM of the
potential function).

In order to improve the convergence of our method with the number $J$, we renormalized the
approximated potential so that the area under it remained
constant whatever value $J$ had. Instead of using $V_i=V(x_i)$ in
Eq.~(\ref{ajbjDef}), we have used $V_i=\alpha V(x_i)$ with $\alpha$
equal to the ratio of the area under $V(x)$ over the area under
$V_{approx}(x)$.

The study of the sech${}^2$ mode profile is a good test for a
numerical method as there exists an analytical expression for
$P_{em}(\kappa_nL)$. In this case we have (see L\"offler
\textit{et~al.}~\cite{Lof97})
\begin{subeqnarray}
\label{PemAnalSech2} t_n^{\pm} & = & \frac{\displaystyle
\Gamma\left[1/2-i(kL+\xi_n^{\pm})\right] \: \cdot \:
  \Gamma\left[1/2-i(kL-\xi_n^{\pm})\right]}{\displaystyle \Gamma\left[-ikL\right] \:
  \cdot \: \Gamma\left[1-ikL\right]} \nonumber \\ & & \\
r_n^{\pm} & = & \frac{\displaystyle \Gamma\left[ikL\right] \:
\cdot \:
  \Gamma\left[1-ikL\right]}{\displaystyle \Gamma\left[1/2+i\xi_n^{\pm}\right]\:
  \cdot \:
  \Gamma\left[1/2-i\xi_n^{\pm}\right]}t_n^{\pm}
\end{subeqnarray}
with $\xi_n^{\pm}=\sqrt{\pm(\kappa_nL)^2-1/4}\,$.

\begin{figure}

\begin{center}
\noindent\fbox{\includegraphics[width=8cm, trim= 30 400 20 120,
keepaspectratio]{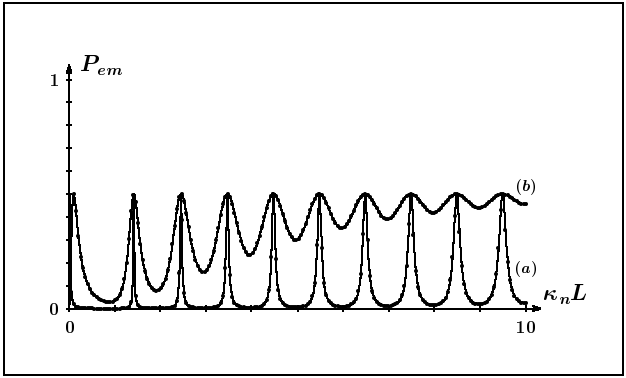}}
\end{center}

\caption{The induced emission probability $P_{em}$ as a function
of
  the interaction length $\kappa_nL$ for $k/\kappa_n=0.01$ (a) and
  $k/\kappa_n=0.1$ (b). The cavity mode is the $\textrm{sech}^2$
  profile. Solid lines were obtained by analytical calculations, dot
  ones by our method ($J=200$).}
  \label{FigurePem_knLSech2}
\end{figure}

Fig.~\ref{FigurePem_knLSech2} shows two $P_{em}(\kappa_nL)$ curves
calculated for $k/\kappa_n=0.01$ and $k/\kappa_n=0.1$ in the case
of the $\textrm{sech}^2$ mode profile. On this figure, the solid
lines represent the analytical results deduced from
formulas~(\ref{TRinExpr}), (\ref{Pem}) and (\ref{PemAnalSech2})
while the dotted ones represent those obtained using our method
(the grid point number $J$ was fixed to 200). The agreement
between these results is very good.

\section{New results}
\label{NewResults}

\subsection{Fundamental sinusoidal mode}

The fundamental sinusoidal mode profile
\begin{equation}
u(x)= \left\{ \begin{array}{ll}\sin(\pi x /L) \quad & \mathrm{for}
\quad 0 < \mathnormal{x} <  L \\ 0 & \mathrm{elsewhere}
\end{array} \right.
\end{equation}
has been studied by L\"offler \textit{et~al.}~\cite{Lof97} and
Retamal \textit{et~al.}~\cite{Ret98}. Their conclusions are not in
agreement with regard to the behaviour of the
induced emission probability for ultracold atoms ($k/ \kappa_n =
0.01$) interacting with large cavities ($\kappa_nL$ of the order
of $10^5$). Such cavity parameters are of a great interest as they
correspond to realistic values for Rydberg atoms interacting with
microwave cavities in recent experiments performed by the Ecole
Normale Sup\'erieure Group \cite{Bru96} (see discussion about the
orders of magnitude in \cite{Ret98}). Retamal
\textit{et~al.}~\cite{Ret98} have predicted well resolved
resonances in the curve $P_{em}(\kappa_nL)$ for the parameters
cited above, whereas the curve predicted on the basis of the results of L\"offler \textit{et~al.}~\cite{Lof97}
looks very different.

\begin{figure}

\begin{center}
\noindent\fbox{\includegraphics[width=8cm, trim= 30 400 20 120,
keepaspectratio]{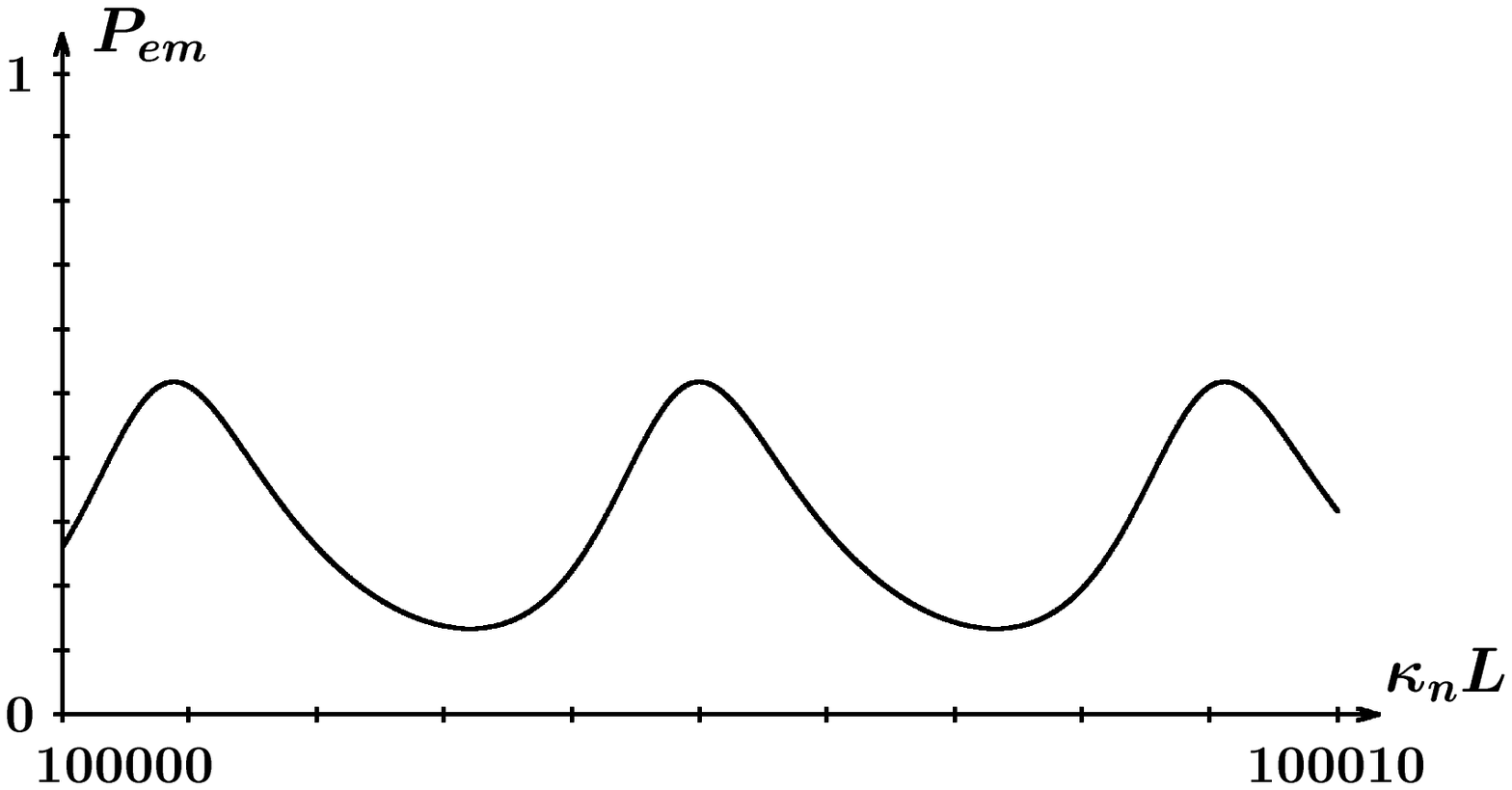}}
\end{center}

\caption{The induced emission probability $P_{em}$ as a function
of the interaction length $\kappa_nL$ for $k/\kappa_n=0.01$
($J=100$, fundamental sinusoidal cavity mode profile).}
  \label{FigurePem_knLFundSin}
\end{figure}

We have calculated the $P_{em}(\kappa_nL)$ curve for an atom
interacting with this mode profile by use of our method.
Fig.~\ref{FigurePem_knLFundSin} presents this curve for
$\kappa_nL$ comprised between 100000 and 100010 ($k/ \kappa_n =
0.01$ and $J=100$). Our curve shows the same well resolved
resonances as those predicted by Retamal
\textit{et~al.} \cite{Ret98}. So we confirm their predictions that
the WKB approximation cannot be used when considering ultracold
atoms with $k/\kappa_n = 0.01$ and realistic interaction lengths of
the order of $10^5$.
Our calculations confirm also that these resonances have
got smeared at $\kappa_nL=10^5$ if we consider warmer atoms
characterized by $k/ \kappa_n = 0.1$. For this last value of $k/ \kappa_n$ the WKB approximation still holds and $P_{em}$ is then invariably
equal to $1/2$. It is good to remark here the importance of still having resolved resonances in $P_{em}$ for actual interaction and cavity parameters, as this result will permit the possibility of testing, in principle, the mazer effect with present cavities. The construction of a re-entrant cavity, as it was proposed in Ref. \cite{Lof97}, would not be necessary for testing the mazer effect, but for improving the resolution of the resonances by approximating a mesa function mode profile.

\subsection{First excited sinusoidal mode}

The first excited sinusoidal mode profile
\begin{equation}
u(x)= \left\{ \begin{array}{ll}\sin(2\pi x /L) \quad &
\mathrm{for} \quad 0 < \mathnormal{x} <  L \\ 0 & \mathrm{elsewhere}
\end{array} \right.
\end{equation}
has been studied by L\"offler \textit{et~al.}~\cite{Lof97}. These
authors have derived an expression for $P_{em}$ on the basis of a
detailed WKB calculation. They have shown that, for large
interaction lengths (without being very explicit about the word
``large''), the behaviour of the induced emission
probability is described by $P_{em}(\kappa_nL)=\sin^2(\Delta_n)$
with
\begin{equation}
  \label{Deltan}
  \Delta_n = \kappa_nL \cdot \frac{1}{\pi}\int_0^{\pi/2}\!\!\!\!\sqrt{(k/\kappa_n)^2 + \cos(x)}dx
\end{equation}
which can be written $\Delta_n = \kappa_nL \cdot Const.$ if $k/\kappa_n$ is a fixed ratio.

\begin{figure}

\begin{center}
\noindent\fbox{\includegraphics[width=8cm, trim= 30 400 20 120,
keepaspectratio]{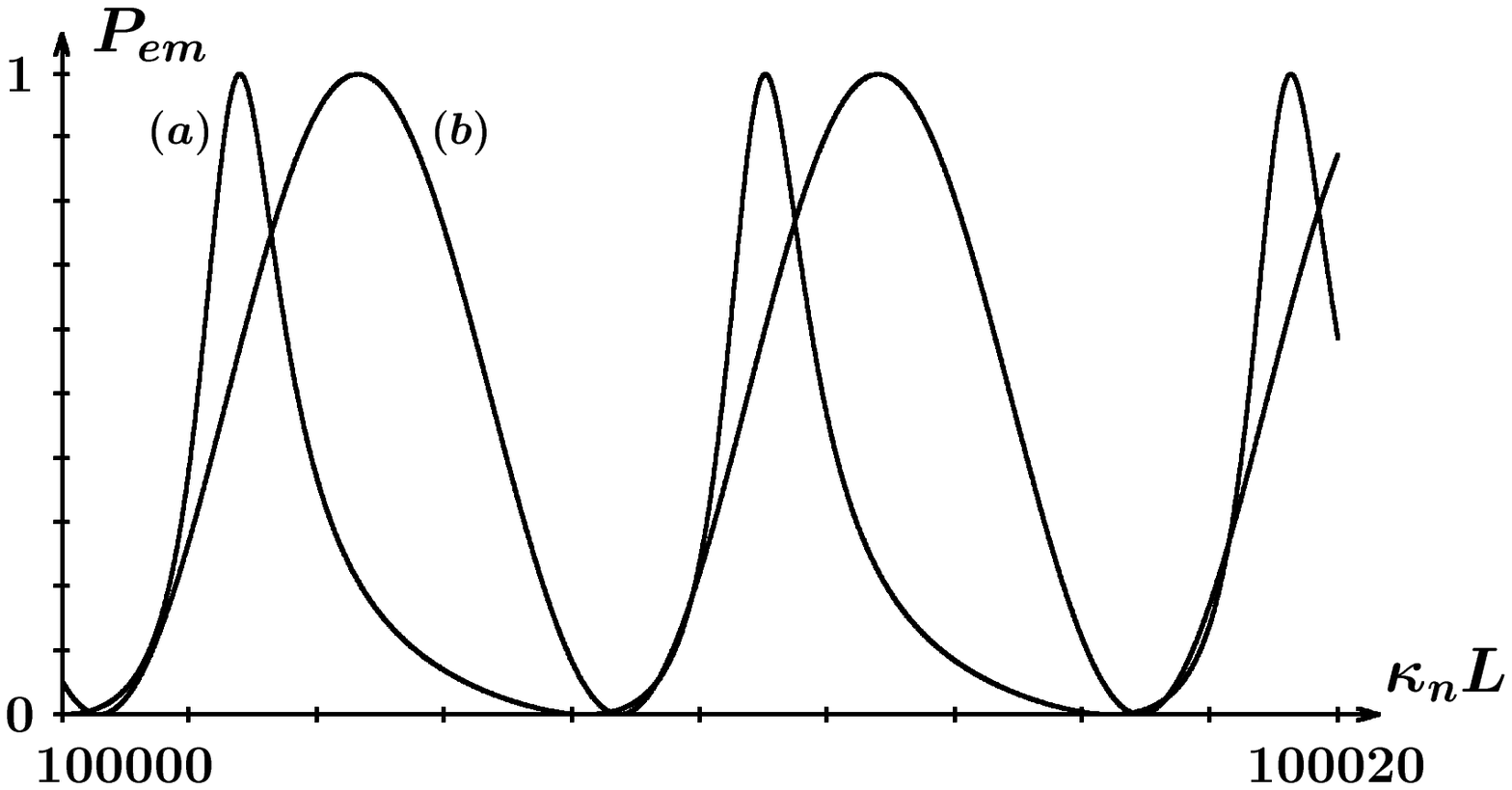}}
\end{center}

\caption{The induced emission probability $P_{em}$ as a function
of
  the interaction length $\kappa_nL$ for $k/\kappa_n=0.01$ (a) and
  $k/\kappa_n=0.1$ (b) ($J=200$,
  first excited sinusoidal cavity mode profile).}
  \label{FigurePem_knLFirstSin}
\end{figure}

In Fig.~\ref{FigurePem_knLFirstSin}, we present two curves of
$P_{em}(\kappa_nL)$ calculated for $\kappa_nL$ comprised between
100000 and 100020 (for $k/\kappa_n = 0.01$ and $k/\kappa_n =
0.1$). The first curve does not exhibit a square sine dependence
over the interaction length, whereas the second one does. This indicates that $\kappa_nL=10^5$ cannot be considered as
``large'' in the sense of L\"offler {\it {et al.}}~\cite{Lof97} for $k/\kappa_n = 0.01$. For $k/\kappa_n = 0.1$, $\Delta_n = 2\pi \kappa_nL / T$, with the period $T$ equals to $\sim\!\!16.3$ (according to
Eq.~(\ref{Deltan})). This period is well reproduced by our
calculations (see curve (b) on Fig.~\ref{FigurePem_knLFirstSin}).

It is a remarkable result of L\"offler {\it {et al.}} \cite{Lof97}
the possibility of building state-changing and state-preserving
mirrors for atoms by modifying the length of the cavity using this
first excited sinusoidal mode. From
Fig.~\ref{FigurePem_knLFirstSin} we conclude that resonances in
$P_{em}$ for the first excited mode are even narrower than those
predicted by a square sine function $\sin^2(\Delta_n)$ . This is a
very convenient result as we are considering actual interaction
and cavity parameters.

\subsection{Gaussian mode}

Up to-date, the Gaussian cavity mode profile has not been studied
exactly in the quantum theory of the mazer. L\"offler
\textit{et~al.}~\cite{Lof97} have argued that the
$\textrm{sech}^2$ mode profile could be used as a good
approximation of the Gaussian one.

To verify this assumption, we have calculated various $P_{em}(\kappa_nL)$ curves for
the profile
\begin{equation}
  \label{gaussianProfile}
  u(x)=e^{-\frac{x^2}{2 \sigma^2}}
\end{equation}

The parameter $\sigma$ was fixed to $\sqrt{2/ \pi} L$ in order to
adopt the same normalization factor for the two profiles
(identical area under the modes).

\begin{figure}

\begin{center}
\noindent\fbox{\includegraphics[width=8cm, trim= 30 400 20 120,
keepaspectratio]{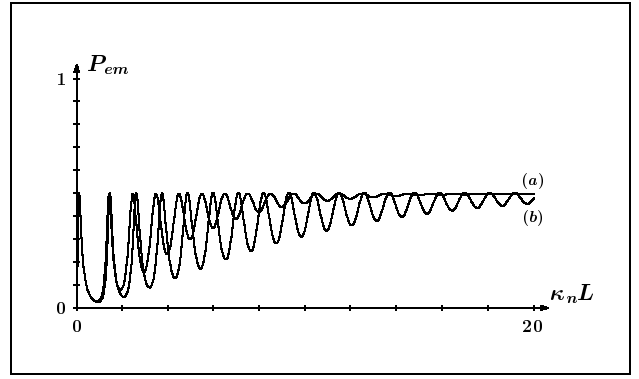}}
\end{center}

\caption{The induced emission probability $P_{em}$ as a function
of the interaction length $\kappa_nL$ for the $\textrm{sech}^2$
mode profile (a) and the Gaussian profile (b) ($k/\kappa_n=0.1$,
$J=300$).} \label{FigurePem_knLGaussian}
\end{figure}

Fig.~{\ref{FigurePem_knLGaussian}} shows our results for
$k/\kappa_n=0.1$ in the range $\kappa_nL=0$ to $\kappa_nL=20$.
Qualitatively both profiles exhibit the same behaviour~: the
resonances in the curves get smeared with increasing values of
$\kappa_nL$. But this phenomenon is not so marked in the case of
the Gaussian profile. Resonances still exist for longer
interaction lengths. This is not a surprising result as the
Gaussian profile is growing more abruptly than the
$\textrm{sech}^2$ one. Thus it is in some sense ``closer'' to the
mesa mode, which exhibits resonances at infinity.

We have also considered the case $k/\kappa_n=0.01$. Our
calculations have shown that 90\% damped oscillations are still
present in the $P_{em}(\kappa_nL)$ curve for interaction lengths
approximately 3 times larger in comparison with the
$\textrm{sech}^2$ mode case. For these calculations, the region $[x_a,x_b]$ was limited to 16 times
the cavity length $L$ and the grid point number $J$ was fixed to 300.

As it was pointed out in Ref. \cite{Lof97} the mazer effect is not
restricted to the microwave domain and it might be tested more
efficiently in the optical domain. Note that typical optical
cavities have a Gaussian mode profile and it is possible to
consider large coupling constants, fact that will help for testing
the mazer effect. We want to call attention to the fact that Hood
{\it et al.}, in Ref. \cite{Hoo98}, presented the first
experimental result for which the interaction energy $\hbar g$ is
greater than the atomic kinetic energy ($k/\kappa_n \ll 1$).

\section{Summary}

We have developed a new method for solving one-dimensional
scattering problems that may be applied advantageously instead of
the WKB approach. This has enabled us to
calculate efficiently the induced emission probability $P_{em}$ of
an atom interacting with a high-$Q$ cavity for various mode
profiles. Two sinusoidal modes have been considered. For these
cases, we have been able to assert that the WKB approximation
cannot be used in the computation of $P_{em}$ for ultracold atoms
($k/ \kappa_n = 0.01$) interacting with actual cavities
characterized by $\kappa_nL$ of the order of $10^5$. Significant
and convenient different physical predictions are found, if we
compare our results with previous works \cite{Lof97}.

The Gaussian mode profile has also been considered and we have
shown that, although it exhibits a similar behaviour in comparison
with the $\textrm{sech}^2$ mode profile, the resonances in the
$P_{em}(\kappa_nL)$ curves exist for significantly larger values
of $\kappa_nL$. The Gaussian mode is relevant when considering
open cavities in the microwave or optical domains.

The presented numerical method will be helpful for computing the
induced emission probability in the case of the recently studied
two-photon mazer \cite{Zha99}, when considering field mode
profiles different from the mesa function. This and other
applications of this numerical method will be presented elsewhere.

\acknowledgements This work has been supported by the Belgian
Institut Interuniversitaire des Sciences Nucl\'eaires (IISN) and
the Brazilian Conselho Nacional de Desenvolvimento Cient\'{\i}fico
(CNPq). E. S. wants to thank Prof. Nicim Zagury for helpful
comments and suggestions.



\clearpage


\renewcommand{\arraystretch}{2}
\begin{table}
  \caption[$f_{j}^{+}(x)$ and $f_{j}^{-}(x)$ functions]{$f_{j}^{+}(x)$ and $f_{j}^{-}(x)$ functions defining the
    general solutions~(\ref{phij}). In this table, $J_{1/3}(x)$ and
    $Y_{1/3}(x)$ denote respectively the first and second kind
    Bessel functions of order 1/3, $I_{1/3}(x)$ and
    $K_{1/3}(x)$ the first and second kind modified Bessel
    functions of order 1/3, $k_j=\sqrt{2(E-V_j)}$,
    $\rho_j=\sqrt{2(V_j-E)}$, and $z_{j}(x)=a_j+b_{j}x$. \\}
    \label{FunctionsTable}
  \begin{center}
  \begin{tabular}{l|l|l}
  $b_j=0$ & $a_j=0$ & $\begin{array}{l} f_{j}^{+}(x)=1 \\
  f_{j}^{-}(x)=x \end{array}$ \\
  \cline{2-3}
    & $a_j>0$ & $\begin{array}{l} f_{j}^{+}(x)=\cos(k_{j}x) \\
  f_{j}^{-}(x)=\sin(k_{j}x) \end{array}$ \\
  \cline{2-3}
    & $a_j<0$ & $\begin{array}{l} f_{j}^{+}(x)=e^{-\rho_{j}x} \\
  f_{j}^{-}(x)=e^{\rho_{j}x} \end{array}$ \\
  \hline
  $b_j \neq 0$ & $z_{j}(x)>0$ & $\begin{array}{l}
  f_{j}^{+}(x)=\sqrt{\left| z_{j}(x) \right|}\:J_{\frac{1}{3}}\!\!\left(
  \frac{2}{3\left| b_j \right|} \left| z_{j}(x) \right|^{\frac{3}{2}}
  \right) \\
  f_{j}^{-}(x)=\sqrt{\left| z_{j}(x) \right|}\:Y_{\frac{1}{3}}\!\!\left(
  \frac{2}{3\left| b_j \right|} \left| z_{j}(x) \right|^{\frac{3}{2}}
  \right) \smallskip \end{array}$ \\
  \cline{2-3}
    & $z_{j}(x)<0$ & $\begin{array}{l}
  f_{j}^{+}(x)=\sqrt{\left| z_{j}(x) \right|}\:I_{\frac{1}{3}}\!\!\left(
  \frac{2}{3\left| b_j \right|} \left| z_{j}(x) \right|^{\frac{3}{2}} \right) \\
  f_{j}^{-}(x)=\sqrt{\left| z_{j}(x) \right|}\:K_{\frac{1}{3}}\!\!\left(
  \frac{2}{3\left| b_j \right|} \left| z_{j}(x) \right|^{\frac{3}{2}}
  \right) \smallskip \end{array}$ \\
  \end{tabular}
  \end{center}
\end{table}
\renewcommand{\arraystretch}{1.0}

\renewcommand{\arraystretch}{1.8}
\begin{table}
  \caption{Behaviour of the $f_{j}^{\pm}(x)$ functions and their
  derivative $g_{j}^{\pm}(x)$ at the classical turning points when $b_j \neq 0$.}
  \label{Limits}
  \begin{center}
  \begin{tabular}{l|l}
    $E > V(x)$ region & $E < V(x)$ region \\
    \hline
    $\begin{array}{l}   \displaystyle f^{+}(x) \rightarrow 0 \\
                        \displaystyle f^{-}(x) \rightarrow -\frac{1}{\pi} |3 b_j|^{1/3} \Gamma(1/3)
    \end{array}$ &
    $\begin{array}{l}   \displaystyle f^{+}(x) \rightarrow 0 \\
                        \displaystyle f^{-}(x) \rightarrow \frac{1}{2} |3 b_j|^{1/3} \Gamma(1/3)
    \end{array}$ \\
    $\begin{array}{l}   \displaystyle g^{+}(x) \rightarrow \frac{b_j}{|3 b_j|^{1/3} \Gamma(4/3)} \\
                        \displaystyle g^{-}(x) \rightarrow \frac{b_j}{\sqrt{3} |3 b_j|^{1/3} \Gamma(4/3)}
    \end{array}$ &
    $\begin{array}{l}   \displaystyle g^{+}(x) \rightarrow -\frac{b_j}{|3 b_j|^{1/3} \Gamma(4/3)} \\
                        \displaystyle g^{-}(x) \rightarrow \frac{\pi b_j}{\sqrt{3} |3 b_j|^{1/3} \Gamma(4/3)}
    \end{array}$ \\
  \end{tabular}
  \end{center}
\end{table}
\renewcommand{\arraystretch}{1.0}

\clearpage


\end{document}